\documentstyle[preprint,epsfig,aps]{revtex}

\begin{document} 

\draft                                                            

\title{The Pauli principle, QRPA and the two-neutrino
double beta decay
\footnote{Supported by the ``Deutsche 
Forschungsgemeinschaft'', contract No. Fa 67/17-1 and
by the ``Graduierten Kolleg - Struktur und Wechselwirkung 
von Hadronen und Kernen'', DFG, Mu 705/3.}}

\author{J. Schwieger, F. \v Simkovic
\thanks{{\it On leave from:}Bogoliubov Theoretical Laboratory, 
Joint Institute for Nuclear Research, 
141980 Dubna, Moscow Region, Russia and Department of Nuclear Physics,  
Comenius University, Mlynsk\'a dolina F1, Bratislava, Slovakia} and
Amand Faessler }

\address{\em Institut f\"ur Theoretische Physik der Universit\"at 
T\"ubingen, \\ Auf der Morgenstelle 14, 72076 T\"ubingen, Germany}

\date{\today}
\maketitle

\begin{abstract}
We examine the violation of the Pauli exclusion principle
in the Quasiparticle Random Phase Approximation (QRPA)
calculation of the two-neutrino double beta decay matrix elements, 
which has its origin in the quasi-boson approximation. For that
purpose we propose a new renormalized QRPA with proton-neutron
pairing method (full-RQRPA) for nuclear structure studies, which 
includes ground state correlation beyond the QRPA. This is achieved 
by using of renormalized quasi-boson approximation, in which the 
Pauli exclusion principle is taken into account more carefully.
The full-RQRPA has been applied to two-neutrino double beta decay
of $^{76}Ge$, $^{82}Se$, $^{128}Te$ and $^{130}Te$. The nuclear matrix
elements have been found significantly less sensitive to the increasing
strength of particle-particle interaction in the physically 
interesting region in comparison with QRPA results. 
The strong differences between the results of both methods indicate that the
Pauli exclusion principle plays an important role in the evaluation
of the double beta decay. The inclusion of the Pauli principle
removes the difficulties with the strong dependence on the 
particle-particle strength $g^{}_{pp}$ in the QRPA on the two-neutrino
double beta decay.
\end{abstract}
\pacs{23.40.Hc}

\section{Introduction}
Neutrinoless double beta decay ($0\nu\beta\beta$), which involves
the emission of two electrons and no neutrinos, demands neutrino
to be a Majorana particle and have a non-zero mass. This process
violates the lepton number conservation and occurs in some theories
beyond the standard model (\cite{1}-\cite{5} for reviews).
But, the $0\nu\beta\beta$-decay has not 
yet been observed. The experimental lower limits on the half-lives
provide the most stringent limits on the effective mass of the 
neutrinos and the parameters of the right-handed currents, 
depending on $0\nu\beta\beta$-decay nuclear matrix elements. 
Two neutrino double beta decay ($2\nu\beta\beta$), which involves 
the emission of two antineutrinos and two electrons, is a second
order weak process fully consistent with the standard model
\cite{1}-\cite{5}. It is
the rarest process observed so far in the nature. Direct 
measurements of $2\nu\beta\beta$-decay half-lives provide a sensitive
test of nuclear structure models. The reliable evaluation of the
$2\nu\beta\beta$-decay nuclear matrix elements is necessary to gain
confidence in the calculated $0\nu\beta\beta$-decay nuclear matrix 
elements.
 
The quasiparticle random phase approximation (QRPA) is the nuclear 
many-body method most widely used to deal with the nuclear structure 
aspects of the double beta decay process \cite{6}-\cite{16}.
The QRPA has been
found successful in explaining the quenching of the 
$0\nu\beta\beta$-decay nuclear matrix elements and bring them into
closer agreement with experimental values.
 But despite this success, 
the QRPA approach to double beta decay has some shortcomings. 
The main problem is that the results are extremely sensitive 
to the renormalization of the particle-particle component of the 
residual interaction, which is in large part responsible 
for suppressing calculated two neutrino decay rate.  This feature 
has been confirmed in the calculations with both schematic contact
$\delta$ interaction and realistic finite range two-body 
interaction. Several alterations of the QRPA have been proposed that
might change that behavior including proton-neutron 
pairing \cite{17,18}, higher order RPA corrections \cite{19} and 
particle number projection \cite{20,21}.
However, none of these extension of the QRPA have changed the rapid
variation of the calculated $2\nu\beta\beta$-decay nuclear matrix 
elements with the increasing strength of the particle-particle force.

The crucial point of the QRPA consists in the so-called quasi-boson 
approximation (QBA) leading to a violation of the Pauli exclusion
principle. The validity of the QBA in the evaluation of the nuclear many 
body Green functions for double beta decay transitions is questioned  
because of the generation of too much ground state correlations 
with increasing strength of the particle-particle interaction leading to 
a collapse of the QRPA. The renormalized RPA \cite{22}-\cite{26}
 being proposed many 
years ago is a method going one step beyond the QBA.
There are many variations of this method \cite{27}-\cite{31},
 which basically consists in
replacing the uncorrelated ground state  (Hartree-Fock
ground state or BCS ground state)
by the correlated ground state ( RPA or QRPA ground state) 
in the calculation of the
expectation value of the commutator of two bifermion operators. 
The topic of this paper is to propose a renormalized QRPA with
proton-neutron pairing and to study the influence of the ground
state correlations beyond the QRPA  on the value of $2\nu\beta\beta$-decay 
matrix elements. 

We note that recently Toivanen and Suhonen have proposed a similar method 
\cite{32}
to study the nuclear double beta decay which however does not take into 
account proton-neutron pairing. Their method has been applied to calculate
the two-neutrino double beta decay of $^{100}Mo$ by considering some
of the ground state correlations beyond the QRPA. 

\section{Formalism} 

We describe the smearing of the Fermi surface by the Hartree- Fock-
Bogoliubov (HFB) method, which includes proton-proton, neutron-neutron
and proton-neutron pairing  \cite{17,18,33}. 
Particle ($c^{+}_{\tau a m_{a}}$ and
$c^{}_{\tau a m_{a}}$, $\tau = p,n$) and quasiparticle 
($a^{+}_{\mu a m_{a}}$ and $a^{}_{\mu a m_{a}}$, $\mu = 1,2$) 
creation and annihilation operators for spherical shell-model
states labeled by ($a~m_a$) are related to each other by the 
Bogoliubov transformation:
\begin{equation}  \left( \matrix{ c^{+}_{p a m_{a} } \cr
c^{+}_{n a m_{a}} \cr c_{p a {\tilde{m}}_{a} } \cr c_{n a {\tilde{m}}_{a}} 
}\right) = \left( \matrix{ 
u_{a 1 p} & u_{a 2 p} & -v_{a 1 p} & -v_{a 2 p} \cr 
u_{a 1 n} & u_{a 2 n} & -v_{a 1 n} & -v_{a 2 n} \cr
v_{a 1 p} & v_{a 2 p} & u_{a 1 p} & u_{a 2 p} \cr  
v_{a 1 n} & v_{a 2 n} & u_{a 1 n} & u_{a 2 n} }\right)
\left( \matrix{ a^{+}_{1 a m_{a}} \cr
a^{+}_{2 a m_{a}} \cr a_{1 a {\tilde{m}}_{a}} \cr a_{2 a {\tilde{m}}_{a}
} }\right),
\label{eq:1}  
\end{equation} 
where the tilde $\sim$ indicates the time reversed states 
$a_{\tau a {\tilde m}_{a}}$ = $(-1)^{j_{a} - m_{a}}a^{}_{\tau a -m_{a}}$. 
The label $a$ designates single particle (quasiparticle) quantum numbers 
($n^{}_a, l^{}_{a}, j^{}_{a}$). The occupation amplitudes $u$ and 
$v$  and the single quasiparticle energies $E^{}_{a \alpha}$ 
are obtained by solving the adequate HFB equation \cite{17,18,33}.  
The vacuum of the quasiparticle operators will be denoted by $|0^+_{HFB}>$.  
We note that in the ground states of spherical even mass nuclei one can 
have only J=0 and T=1 proton-neutron pairs in the framework of the BCS
and more general HFB theory.
In principle one could have T=0, J=odd correlations but only for deformed
nuclei \cite{33a}. The treatment here is restricted to spherical nuclei.
Such deformation effects are taken effectively into account by 
renormalizing according to \cite{17,18} the T=1, $J=0^+$ proton-neutron
pairing to be stronger than requested by the Brueckner reaction matrix 
elements of the Bonn potential. This strength is fitted to the 
experimental proton-neutron pairing gap extracted from the masses
\cite{17,18}.

In the limit in which there is no proton-neutron pairing
$u^{}_{2p}$ = $v^{}_{2p}$ = $u^{}_{1n}$ = $v^{}_{1n}$ = 0. Then 
the HFB transformation in Eq. (\ref{eq:1}) reduces  to two 
BCS transformations, first for protons ($u^{}_{1p}=u^{}_p$, 
$v^{}_{1p}=v^{}_p$) and second for neutrons 
($u^{}_{2n}=u^{}_n$, $v^{}_{2n}=v^{}_n$).

We assume the nuclear motion to be harmonic and the phonon operator
to be of the form \cite{17,18,33}
\begin{eqnarray}
Q^{m+}_{JM^\pi} = \sum_{k l} 
\{ X^{m}_{12}(k,l,J^{\pi})  
A^{+}_{12}(k,l,J,M) + Y^{m}_{12}(k,l,J^{\pi}) 
{\tilde A}_{12}(k,l,J,M) \}
\nonumber \\
+ \sum_{k\leq l \atop \mu = 1,2}
\{ X^{m}_{\mu \mu}(k,l,J^{\pi})  
A^{+}_{\mu \mu}(k,l,J,M) + Y^{m}_{\mu \mu}(k,l,J^{\pi})  
{\tilde A}^{}_{\mu \mu}(k,l,J,M) \},
\label{eq:2}    
\end{eqnarray}
Here, $X^m_{}$ and $Y^m_{}$ are forward-
and backward- going free variational amplitudes. The operator 
$A^+_{\mu \nu}(k,l,J,M)$ ($\mu, \nu = 1,2$) creates 
a pair of quasiparticles coupled to angular momentum J with projection 
M, namely 
\begin{eqnarray}
A^+_{\mu \nu}(k, l, J, M) &=& n(k\mu, l\nu) \sum^{}_{m_k , m_l }
C^{J M}_{j_k m_k j_l m_l } a^+_{\mu k m_k} a^+_{\nu l m_l},
\nonumber \\
{\tilde A}^{}_{\mu \nu}
(k, l, J M) &=& (-1)^{J-M} A^{}_{\mu \nu}(k, l, J, -M)
\label{eq:3}    
\end{eqnarray}
The factor $n(k \mu , l \nu)$ is a normalization in the case in which 
both quasiparticles are in the same shell: 
\begin{equation}
n(k\mu, l \nu)=
(1+(-1)^J\delta_{kl}\delta_{\mu \nu})/(1+\delta_{kl}\delta_{\mu
\nu})^{3/2}.
\label{eq:4}  
\end{equation}

By using the machinery of the equation of motion 
we get eigenvalue equation
\begin{equation}
\label{qrpa_eq}
\left(
\begin{array}{cc}
\cal A&\cal B\\
\cal B&\cal A
\end{array}
\right)_{J^\pi}
\left(
\begin{array}{c}
X^m\\
Y^m
\end{array}
\right)_{J^\pi}
= \Omega^m_{J^\pi}
\left( 
\begin{array}{cc}
\cal U&0\\
0&\cal -U
\end{array}
\right)_{J^\pi}
\left(
\begin{array}{c}
X^m\\
Y^m
\end{array}
\right)_{J^\pi}
,
\label{eq:5}  
\end{equation}
with the matrices
\begin{equation}
{\cal A}^{}_{J}(\mu k, \nu l; \mu' k', \nu' l')
=\big<0^+_{RPA}\big|\big
[A^{}_{\mu  \nu}(k, l, J, M),
[\hat{H},A^+_{\mu' \nu'}(k', l', J, M)]\big]\big|0^+_{RPA}\big>, 
\label{eq:6}  
\end{equation} 
\begin{equation}
{\cal B}^{}_{J}(\mu k, \nu l; \mu' k', \nu' l')
=\big<0^+_{RPA}\big|\big
[A^{}_{\mu \nu}(k, l, J, M),
[\tilde{A}^{}_{\mu' \nu'}(k', l', J, M),\hat{H}]\big]\big|0^+_{RPA}\big>,
\label{eq:7}  
\end{equation}
and the unitary-matrix $\cal U$ 
\begin{equation}
{\cal U}=\big<0^+_{RPA}\big|[A^{}_{\mu \nu}(k, l, J, M ),
 A^+_{\mu' \nu'}(k', l', J, M)]\big|0^+_{RPA}\big>.
\label{eq:8}  
\end{equation}
Here, $\hat{H}$ is the quasiparticle hamiltonian and 
$\Omega^m_{J^\pi}$ is the excitation energy $E_m^{J^\pi}-E_0$. 
$E_m^{J^\pi}$ and $E_0$ are respectively eigenenergies of the 
excited state $|JM^{\pi}_{m}>$ 
($|JM^{\pi}_{m}>$ $=Q^{m+}_{JM^\pi}|0^+_{RPA}>$) and of the 
ground state $|0^+_{RPA}>$ ($Q^{m}_{JM^\pi}|0^+_{RPA}>=0)$.

In order to solve the Eq. (\ref{eq:5})  we need additional 
approximation. In the framework of QBA
\begin{eqnarray}
\lefteqn{
\big[A^{}_{\mu \nu}(k, l, J, M),A^+_{\mu' \nu'}(k', l', J, M)\big] }
\nonumber \\
&&\simeq
\big<0^+_{HFB}\big|\big
[A^{}_{\mu \nu}(k, l, J, M),A^+_{\mu' \nu'}(k', l', J, M)\big]
\big|0^+_{HFB}\big> 
\nonumber \\ 
&&=n(k\mu, l\nu) n(k'\mu', l'\nu') 
\Big( \delta_{kk'}\delta_{\mu \mu' }\delta_{ll'}
\delta_{\nu\nu'} -
\delta_{lk'}\delta_{\nu \mu'}\delta_{kl'}
\delta_{\mu \nu'}(-1)^{j_{k}+j_{l}-J}\Big), 
\label{eq:9}  
\end{eqnarray}
the unitary-matrix $\cal U$ in Eq. (\ref{eq:5})  is 
just unity-matrix and the matrix equation (\ref{eq:5}) 
simplifies to QRPA with proton-neutron pairing
equation (full-QRPA) \cite{17,18,33}.  

It is well-known that the QBA violates the Pauli exclusion 
principle because we have neglected terms coming from the
commutator. This deficiency could be improved with help
of renormalized quasi-boson approximation (RQBA)
\begin{eqnarray}
\lefteqn{
\big[A^{}_{\mu \nu}(k, l, J, M),A^+_{\mu' \nu'}(k', l', J, M)\big] }
&&\nonumber \\ 
&&\simeq
\big<0^+_{RPA}\big|\big
[A^{}_{\mu \nu}(k, l, J, M),A^+_{\mu' \nu'}(k', l', J, M)\big]
\big|0^+_{RPA}\big> 
\nonumber \\ 
&&=n(k\mu, l\nu) n(k'\mu', l'\nu') 
\Big( \delta_{kk'}\delta_{\mu \mu' }\delta_{ll'}
\delta_{\nu\nu'} -
\delta_{lk'}\delta_{\nu \mu'}\delta_{kl'}
\delta_{\mu \nu'}(-1)^{j_{k}+j_{l}-J}\Big) \nonumber \\
&&\times \underbrace{
\Big\{1
\,-\,\frac{1}{\hat{\jmath}_{l}}
<0^+_{RPA}|[a^+_{\nu l}a_{\nu \tilde{l}}]_{00}|0^+_{RPA}>
\,-\,\frac{1}{\hat{\jmath}_{k}}
<0^+_{RPA}|[a^+_{\mu k}a_{\mu \tilde{k}}]_{00}|0^+_{RPA}>
\Big\}
}_{
=:\displaystyle {\cal D}_{\mu k, \nu  l; J^\pi}
},
\label{eq:10}  
\end{eqnarray}
with the notations $\hat{\jmath}_k=\sqrt{2j_k+1}$ and 
\begin{equation}
[a^+_{\mu k}a_{\mu \tilde{k}}]_{00}
=\sum_{m_{k}}C^{00}_{j_k m_k j_{k} -m_{k}}
a^+_{\mu k m_k} a_{\mu k -m_{k}}.
\label{eq:11}  
\end{equation}
If ${\cal D}_{\mu k, \nu  l; J^\pi} = 1$, one recovers the usual boson
commutation relations for the QBA. The reduction of 
${\cal D}_{\mu k, \nu  l; J^\pi}$ from 1 reflects the fact that one has 
the commutation relations of Fermion pairs (Pauli principle) and not of
bosons. 

It is useful to introduce renormalized operators
\begin{equation}
\overline{A}^{+}_{\mu \nu}(k, l, J, M)~=~
{\cal D}^{-1/2}_{\mu k, \nu  l; J^\pi}
~A^{+}_{\mu \nu}(k, l, J, M).
\label{eq:12}  
\end{equation}
In the RQBA the overlap matrix $\cal U$ is also diagonal: 
${\cal U} = D^{1/2}_{} 
{\scriptsize\left(\begin{array}{cc}{1}&0\\0&{1}\end{array}\right)}
D^{1/2}_{}$. After the transformations
\begin{eqnarray}
\overline{\cal A}&=&{\cal D}^{-1/2}{\cal A}{\cal D}^{-1/2},~~~~~~
\overline{\cal B}={\cal D}^{-1/2}{\cal B}{\cal D}^{-1/2},\nonumber \\
\overline{X}^m&=&{\cal D}^{1/2}X^m,~~~~~~~
\overline{Y}^m={\cal D}^{1/2}Y^m,
\label{eq:13}  
\end{eqnarray}
for the phonon operator we can write
\begin{eqnarray}
Q^{m+}_{JM^\pi} = \sum_{\mu k \leq \nu l} 
\{ \overline{X}^{m}_{\mu \nu}(k,l,J^{\pi})  
\overline{A}^{+}_{\mu \nu}(k,l,J,M) + 
\overline{Y}^{m}_{\mu \nu}(k,l,J^{\pi}) 
{\tilde {\overline{A}}}_{\mu \nu}(k,l,J,M) \}
\label{eq:14}    
\end{eqnarray}
and the renormalized quasiparticle random phase approximation (full-RQRPA)
equation takes the form of RPA-like equation
\begin{equation}
\left(\begin{array}{cc}
\overline{\cal A}&\overline{\cal B}\\
\overline{\cal B}&\overline{\cal A}\end{array}\right)_{J^\pi}
\left(
\begin{array}{c}\overline{X}^m\\ \overline{Y}^m \end{array}
\right)_{J^\pi}
= \Omega^m_{J^\pi}
\left(
\begin{array}{cc}
1&0\\
0&-1
\end{array}
\right)
\left(
\begin{array}{c}
\overline{X}^m\\
\overline{Y}^m
\end{array}
\right)_{J^\pi}.
\label{eq:15}  
\end{equation}
The matrix elements $\overline{\cal A}$ and $\overline{\cal B}$ are explicitly
given by 
\begin{eqnarray}
\overline{\cal A}^{}_{J^\pi}
(\mu k, \nu l; \mu' k', \nu' l')  
\nonumber
\end{eqnarray}
\begin{eqnarray}
= n(k\mu,l\nu)n(\acute{k}\acute{\mu},\acute{l}\acute{\nu}) 
\{~ 
(E_{k\mu}+E_{l\nu})(\delta_{k\acute{k}}\delta_{\mu\acute{\mu}}
\delta_{l\acute{l}}\delta_{\nu\acute{\nu}}-(-1)^{j_{k}+j_{l}-J}
\delta_{k\acute{l}}\delta_{\mu\acute{\nu}}
\delta_{l\acute{k}}\delta_{\nu\acute{\mu}})  \nonumber \\ 
- 2~{\cal D}^{1/2}_{\mu k, \nu  l; J^\pi}~
{\cal D}^{1/2}_{\mu' k', \nu'  l'; J^\pi}
\nonumber \\
\times\sum_{\alpha\beta\delta\gamma}[~ 
G(k\alpha l\beta \acute{k}\gamma \acute{l}\delta J )
(u_{k\mu\alpha}u_{l\nu\beta}u_{\acute{k}\acute{\mu}\gamma}
u_{\acute{l}\acute{\nu}\delta}+v_{k\mu\alpha}v_{l\nu\beta}
v_{\acute{k}\acute{\mu}\gamma}v_{\acute{l}\acute{\nu}\delta})
\nonumber \\ 
+F(k\alpha l\beta \acute{k}\gamma \acute{l}\delta J )
(u_{k\mu\alpha}v_{l\nu\beta}u_{\acute{k}\acute{\mu}\gamma}
v_{\acute{l}\acute{\nu}\delta}+v_{k\mu\alpha}u_{l\nu\beta}
v_{\acute{k}\acute{\mu}\gamma}u_{\acute{l}\acute{\nu}\delta})
\nonumber \\
-(-1)^{j_{\acute{k}}+j_{\acute{l}}-J}
F(k\alpha l\beta \acute{l}\delta \acute{k}\gamma J )
(u_{k\mu\alpha}v_{l\nu\beta}u_{\acute{l}\acute{\nu}\delta}
v_{\acute{k}\acute{\mu}\gamma}+v_{k\mu\alpha}u_{l\nu\beta}
v_{\acute{l}\acute{\nu}\delta}u_{\acute{k}\acute{\mu}\gamma})
~]~~\}, 
\label{eq:16}  
\end{eqnarray}

\begin{eqnarray}
\overline{\cal B}^{}_{J^\pi}(\mu k, \nu l; \mu' k', \nu' l')  
~~~~~~~~~=~2~n(k\mu,l\nu)~n(\acute{k}\acute{\mu},\acute{l}\acute{\nu})
~{\cal D}^{1/2}_{\mu k, \nu  l; J^\pi}~
{\cal D}^{1/2}_{\mu' k', \nu'  l'; J^\pi}
\nonumber \\
\times\{
~~\sum_{\alpha\beta\delta\gamma}[~ 
G(k\alpha l\beta \acute{k}\gamma \acute{l}\delta J )
(u_{k\mu\alpha}u_{l\nu\beta}v_{\acute{k}\acute{\mu}\gamma}
v_{\acute{l}\acute{\nu}\delta}+v_{k\mu\alpha}v_{l\nu\beta}
u_{\acute{k}\acute{\mu}\gamma}u_{\acute{l}\acute{\nu}\delta})
\nonumber \\ 
-F(k\alpha l\beta \acute{k}\gamma \acute{l}\delta J )
(u_{k\mu\alpha}v_{l\nu\beta}v_{\acute{k}\acute{\mu}\gamma}
u_{\acute{l}\acute{\nu}\delta}+v_{k\mu\alpha}u_{l\nu\beta}
u_{\acute{k}\acute{\mu}\gamma}v_{\acute{l}\acute{\nu}\delta})
\nonumber \\
+(-1)^{j_{\acute{k}}+j_{\acute{l}}-J}
F(k\alpha l\beta \acute{l}\delta \acute{k}\gamma J )
(u_{k\mu\alpha}v_{l\nu\beta}v_{\acute{l}\acute{\nu}\delta}
u_{\acute{k}\acute{\mu}\gamma}+v_{k\mu\alpha}u_{l\nu\beta}
u_{\acute{l}\acute{\nu}\delta}v_{\acute{k}\acute{\mu}\gamma})
~]~~\}.
\label{eq:17}  
\end{eqnarray}
Here, $G(a\alpha b \beta c \gamma d \delta J)$ and $F(a\alpha b \beta c
\gamma d \delta J)$ are respectively particle-particle and particle-hole
interaction matrix-elements defined in \cite{34}.  The greek 
letters $\alpha$,
$\beta$, $\gamma$ and $\delta$ are isospin indices 
(e.g., $\alpha = p, n$) while the greek letters
$\mu$, $\mu'$, $\nu$ and $\nu'$ run the value $1$ or $2$ and
denote the ``quasi-isospin'' of the quasiparticles.

In order to calculate the RQRPA matrices $\overline{\cal A}^{}_{J^\pi}$
and $\overline{\cal B}^{}_{J^\pi}$ one has to determine quantity 
${\cal D}_{\mu k, \nu  l; J^\pi}$. We follow the method of ref. \cite{31}
and express the one-body densities of Eq. (\ref{eq:10}) in terms 
of mappings 
\begin{equation}
[a^+_{\mu k}a_{\mu \tilde{k}}]_{00} = 
\frac{1}{{\hat{j}}^{}_{k}}\sum_{\nu l \atop J M}
A^+_{\mu \nu}(k, l, J, M) A^{}_{\mu \nu}(k, l, J, M). 
\label{eq:18}  
\end{equation}
By using eqs. (\ref{eq:10}), (\ref{eq:12}), (\ref{eq:13}),
 (\ref{eq:18}) together with   
\begin{equation}
\overline{A}^{+}_{\mu \nu}(k, l, J, M) = \sum_{m} \{ 
\overline{X}^{m}_{\mu \nu}(k,l,J^{\pi})Q^{m+}_{JM^\pi} -
\overline{Y}^{m}_{\mu \nu}(k,l,J^{\pi}){\tilde Q}^{m}_{JM^\pi} \},
\label{eq:19}  
\end{equation}
we obtain the following expression for ${\cal D}_{\mu k, \nu  l; J^\pi}$:
\begin{eqnarray}
{\cal D}_{k\mu l\nu J^\pi} &=& 1-\frac{1}{\hat{\jmath}_k^2} 
\sum_{k'\mu' \atop J'^{\pi'} m}
{\cal D}_{k\mu k'\mu'J'^{\pi'}}\hat{J}'^2 \big|
\overline{Y}^m_{\mu \mu'}(k, k', J'^{\pi'})  \big|^2
\nonumber \\
&& ~~~~~-\frac{1}{\hat{\jmath}_l^2}\sum_{l'\nu' \atop J'^{\pi'} m}
{\cal D}_{l\nu
l'\nu'J'^{\pi'}}\hat{J}'^2 \big |\overline{Y}^m_{\nu \nu'}(l, l', J'^{\pi'})
  \big|^2
\label{eq:20}  
\end{eqnarray}
The system of non-linear equations in Eq. (\ref{eq:20}) 
could be solved numerically by an iteration process. 
As input the solution of the full-RQRPA equations for all the 
considered multipolarities  is required. However, in order to calculate 
the full-RQRPA equation in Eq. (\ref{eq:15}) we need the knowledge of the
renormalization factors ${\cal D}_{k\mu l\nu J^\pi} $.
Therefore, the selfconsistent scheme of the calculation is a doubly 
iterative problem.  As starting values for ${\cal D}_{k\mu l\nu J^\pi}$
for the first iteration could be used e.g.,
\begin{equation}
{\cal D}_{k\mu l\nu J^\pi} = 1-\frac{1}{\hat{\jmath}_k^2} 
\sum_{k'\mu' \atop J'^{\pi'} m}
\hat{J}'^2 \big|{Y}^m_{\mu \mu'}(k, k', J'^{\pi'})  \big|^2
-\frac{1}{\hat{\jmath}_l^2}\sum_{l'\nu' \atop J'^{\pi'} m}
\hat{J}'^2 \big |{Y}^m_{\nu \nu'}(l, l', J'^{\pi'})  \big|^2,
\label{eq:21}  
\end{equation}
with the amplitudes $Y^{m}$ being the solution of the full-QRPA equation.

The half-life for $2\nu\beta\beta$-decay could be written in a 
factorized form \cite{1}-\cite{5}:
\begin{equation}
\big(T^{2\nu\beta\beta}_{1/2}\big) ^{-1}_{}~ = 
~G^{2\nu}_{GT}~\big| M^{2\nu}_{GT} \big|^2,
\label{eq:22}  
\end{equation}
where $G^{2\nu}_{GT}$ is a lepton phase space integral and the 
nuclear matrix elements $M^{}_{GT}$ in the full-RQRPA is
\begin{equation}
M^{2\nu}_{GT}=\sum_{m^{}_{i},m^{}_{f}}
\frac{^{}_{f}<0_{RPA}^+||\tau^+\sigma||1_{m^{}_{f}}^+>
<1_{m^{}_{f}}^+ | 1_{m^{}_{i}}^+> 
<1_{m^{}_{i}}^+ ||\tau^+\sigma||0_{RPA}^+>_{i}^{}}
{\Omega^{m^{}_{f}}_{1^+}+\Omega^{m^{}_{i}}_{1^+}}
\label{eq:23}  
\end{equation}
with reduced transition-amplitudes of the Gamow-Teller beta decay operator
\begin{eqnarray}
\lefteqn{
<1_{m^{}_{i}}^+ ||\tau^+\sigma||0_{RPA}^+>_{i} ~=~\sum_{\mu k \leq \nu l}~
\sigma(k,l)~({\cal D}^{i}_{k\mu l\nu 1^+})^{1/2} }&&  \nonumber \\
&&\times~\big[~ 
\overline{X}^{m^{}_{i}}_{\mu \nu}(k, l, 1^+)
\big( u^{i}_{k \mu p} v^{i}_{l \nu n} - v^{i}_{k \mu n} u^{i}_{l \nu p} 
\big)~
+~\overline{Y}^{m^{}_{i}}_{\mu \nu}(k, l, 1^+)
\big( v^{i}_{k \mu p} u^{i}_{l \nu n} - u^{i}_{k \mu n} v^{i}_{l \nu p} 
\big) ~\big],
\label{eq:24}  
\end{eqnarray}
\begin{eqnarray}
\lefteqn{
^{}_{f}<0_{RPA}^+||\tau^+\sigma|| 1_{m^{}_{f}}^+>
~=~\sum_{\mu k \leq \nu l}~\sigma(k,l) 
~({\cal D}^{f}_{k\mu l\nu 1^+})^{1/2} }&&  \nonumber \\
&&\times \big[~
\overline{X}^{m^{}_{f}}_{\mu \nu}(k, l, 1^+)
\big( v^{f}_{k \mu p} u^{f}_{l \nu n} - u^{f}_{k \mu n} v^{f}_{l \nu p} 
\big)~
+~\overline{Y}^{m^{}_{f}}_{\mu \nu}(k, l, 1^+)
\big( u^{f}_{k \mu p} v^{f}_{l \nu n} - v^{f}_{k \mu n} u^{f}_{l \nu p} 
\big) ~\big]. 
\label{eq:25}  
\end{eqnarray}
Here, $\sigma (k,l)=<k||\sigma ||l>$ is the reduced matrix element 
of $\sigma$ operator.
The HFB $u$ and $v$ factors entering in eqs. (\ref{eq:24})
and (\ref{eq:25}) belong to two different even-even system. The 
occupation amplitudes $ u^{i}, v^{i}$ and  $ u^{f}, v^{f}$  are obtained
by solving the HFB equation for the initial and final nuclear states, 
respectively.
$|1_{m^{}_{i}}^+>$ and $|1_{m^{}_{f}}^+>$  
are respectively the excited states calculated from the ground state of 
initial (A,Z) and final (A,Z+2) nuclei with the excitation energies 
$\Omega^{m^{}_{f}}_{1^+}$ and  $\Omega^{m^{}_{i}}_{1^+}$.
The states $|1_{m^{}_{i}}^+>$ 
($|1_{m^{}_{f}}^+>$) are expressed in terms of forwards-
$\overline{X}^{m_{i}^{}}_{\mu \nu}(k, l, 1^+)$ 
($\overline{X}^{m_{f}^{}}_{\mu \nu}(k, l, 1^+)$)
and backwards-  $\overline{Y}^{m_{i}^{}}_{\mu \nu}(k, l, 1^+)$ 
($\overline{Y}^{m_{f}^{}}_{\mu \nu}(k, l, 1^+)$) going amplitudes. 
Unfortunately the states $|m_{i}, 1_{}^+>$ and $|m_{f}, 1_{}^+>$ 
are not orthogonal to each other. 
For the overlap matrix of these states we write
\begin{equation}
<1_{m^{}_{f}}^+ | 1_{m^{}_{i}}^+> =
\sum_{\mu k \leq \nu l}~\big(
\overline{X}^{m_{i}^{}}_{\mu \nu}(k, l, 1^+)
\overline{X}^{m_{f}^{}}_{\mu \nu}(k, l, 1^+)-
\overline{Y}^{m_{i}^{}}_{\mu \nu}(k, l, 1^+)
\overline{Y}^{m_{f}^{}}_{\mu \nu}(k, l, 1^+) \big).
\label{eq:26}  
\end{equation}

We note that in the limit of small ground state correlations 
${\cal D}_{k\mu l\nu J^\pi} \simeq 1$ the full-RQRPA
could be replaced by the full-QRPA. The deviation between the results
of both methods is a signal showing on the restriction of the validity of 
the QBA in a given physical region. 
Further, in the limit in which there is no proton-neutron
paring, the full-QRPA reduces to two matrix equations \cite{17,18,33}: 
The first
pp+nn QRPA equation has the origin in the proton-proton and neutron-neutron
quasiparticle excitations. The second pn-QRPA  equation considers only the
proton-neutron quasiparticle excitations from the ground state. 
By neglecting the proton-neutron pairing interaction 
the full-RQRPA equation decomposes in a similar way on pp+nn RQRPA
and pn-RQRPA equations. However, pp+nn RQRPA and pn-RQRPA equations
remain coupled through the Eq. (\ref{eq:20}). Therefore, 
both equations have to be solved parallel in the iterative process.

\section{Calculation and discussion of the results}
We applied the full-RQRPA method to the $2\nu\beta\beta$-decay
of $^{76}Ge$, $^{82}Se$, $^{128}Te$ and $^{130}Te$. We assumed 
the single particle model space both for protons and neutrons as follows.

(i) For $^{76}Ge \rightarrow ^{76}Se$ and $^{82}Se \rightarrow ^{82}Kr$ 
the model space comprises 13 levels:
\begin{displaymath}
1s^{}_{1/2}, 0d^{}_{5/2}, 0d^{}_{3/2},~ 1p^{}_{3/2}, 1p^{}_{1/2},
 0f^{}_{7/2}, 0f^{}_{5/2},~2s^{}_{1/2}, 1d^{}_{5/2}, 1d^{}_{3/2},
 0g^{}_{9/2}, 0g^{}_{7/2}, ~0h^{}_{11/2}. 
\end{displaymath}

(i) For $^{128}Te \rightarrow ^{128}Xe$ and 
$^{130}Te \rightarrow ^{130}Xe$ we used 16 levels:
\begin{displaymath}
 1p^{}_{3/2}, 1p^{}_{1/2},
 0f^{}_{7/2}, 0f^{}_{5/2},~2s^{}_{1/2}, 1d^{}_{5/2}, 1d^{}_{3/2},
 0g^{}_{9/2}, 0g^{}_{7/2}, ~0h^{}_{11/2}, 
 0h^{}_{9/2}, 1f^{}_{7/2}, 1f^{}_{5/2}, 
\end{displaymath}
\begin{displaymath}
 2p^{}_{3/2}, 2p^{}_{1/2},~
 0i^{}_{13/2}.
\end{displaymath}

The single particle energies have been calculated with a 
Coulomb-corrected Wood-Saxon potential. For the two body interaction
we used the nuclear G-matrix calculated from Bonn one-boson 
exchange potential. The single quasiparticle energies and 
occupation amplitudes have been found by solving the HFB equation with
p-n pairing  for both  the parent and the daughter nuclei
in the above mentioned space. The renormalization of the 
proton-proton, neutron-neutron and proton-neutron pairing interaction
has been determined according to ref. \cite{17,18}. The renormalization
of the T=1 J=0 proton-neutron force is due to the quadrupole 
deformation of the intrinsic nucleus. It could also be described 
in a spherical basis by admixing to the single particle propagator
$2^+$ quadrupole excitations. Thus this renormalization and the 
HFB step influence  only the quasiparticle states. The collective 
excitations are described in a second QRPA step.
The renormalization we fit to the proton-neutron pairing gap
\cite{17,18}.

In the calculation of the full-RQRPA equation we renormalized 
particle- particle and particle- hole channels of the G-matrix 
interaction by introducing two parameters $g^{}_{pp}$ and $g^{}_{ph}$,
which in principle should be equal to unity. Since the 
matrix element $M^{2\nu}_{GT}$ is much more sensitive to $g^{}_{pp}$
than to $g^{}_{ph}$, we will set $g^{}_{ph}=1.0$ and discuss the
dependence on $g^{}_{pp}$. Only for the $2^+$-channel of the 
interaction $g^{}_{ph}$ was fixed to $g^{}_{ph}=0.8$ for all studied 
$2\nu\beta\beta$-decay transitions. For higher value of $g^{}_{ph}$ the
 particle-hole interaction in $2^+$ channel is too strong. 
The lowest eigenvalue becomes the imaginary and leads to a collapse of the
correlated ground state. 

It is worthwhile mentioning that the calculation of $M^{2\nu}_{GT}$ within
full-RQRPA needs a great computational effort. We remind that the
full-RQRPA self- consistent  scheme of the calculation requires the
solution of the full-RQRPA equation (\ref{eq:15}) for all 
multipolarities $J^{\pi}_{}$ in each iteration for the initial 
and the final nuclei. In comparison 
with the full-RQRPA, the full-QRPA calculation of $M^{}_{GT}$ 
requires to solve the full-QRPA equation only  for the 
multipolarity $1^+$  once for initial and once for final nucleus. 
The iterative procedure of the full-RQRPA have been found 
to converge rapidly.

As the full-RQRPA calculation is a very time consuming process, it is 
interesting to determine, which multipolarities $J^{\pi}$ are 
important to take into account to calculate the renormalization
factors  ${\cal D}_{\mu k, \nu  l; J^\pi}$ in Eq. (\ref{eq:20}).
We made a calculation for $g^{}_{pp}=1.0$ including more and more
multipolarities $J^{\pi}$ in the renormalization scheme  of the
full-RQRPA. The results are shown in Fig. 1. Reading from the left to the
right, the quoted multipolarity is added at every point to the ones
of the preceding point. One can clearly see a
saturation effect for large multipolarities. This can be explained
by the fact that the higher multipolarities are less collective 
and do only little contribute to the ground state correlations.
Practically, one is able to find
a combination of the multipolarities, that optimizes both the effect of
renormalization and the effort of the computing-time.
Fortunately, it turns out that one combination, which consists of
the $1^+$, $2^+$, $2^-$ and $3^-$ multipolarities, 
shows as good results as the whole set. The corresponding values 
of $M^{2\nu}_{GT}$ for the studied nuclear transitions
are plotted with empty symbols in Fig. 1. A given combination
of the multipolarities have been used to study the $g^{}_{pp}$
dependence of the matrix element $M^{2\nu}_{GT}$.

The goal of the present paper is to study the reliability of the
QBA in the nuclear structure calculation of $M^{2\nu}_{GT}$.
This information can be obtained by the comparison of the results
of the full-RQRPA and the full-QRPA in the physically acceptable
region of the $g^{}_{pp}$ parameters $0.8 \leq g^{}_{pp} \leq 1.2$.  
Our results for $2\nu\beta\beta$-decay of $^{76}Ge$, $^{82}Se$,
$^{128}Te$ and $^{130}Te$ are presented in Fig. 2, Fig. 3, Fig. 4 
and Fig. 5, respectively. We see that the effect due to the
ground state correlations beyond the RPA is improving the agreement for all
studied nuclear transitions. The inclusion of the ground state 
correlations beyond RPA was found to stabilize the behaviour of 
$M^{2\nu}_{GT}$ as a function of $g^{}_{pp}$. 
The full-QRPA and the full-RQRPA equations are of the same form,
differing only by a renormalization of the two-body interaction
by the factor ${\cal D}_{\mu k, \nu  l; J^\pi}$ [see Eq.  (\ref{eq:15})].
For ${\cal D}_{\mu k, \nu  l; J^\pi} \simeq 1$, what is the case
for small ground state correlations, the results of both methods 
are the same. However,  
the QRPA quenching mechanism for $M^{2\nu}_{GT}$ has its origin in 
generating  too much ground state correlations with increasing
strength of the particle-particle interaction.  Strong ground 
state correlations lead to a significantly smaller value of 
${\cal D}_{\mu k, \nu  l; J^\pi}$ and reduces the strength of
two-body interaction. In this way a damping of the 
dependence of $M^{2\nu}_{GT}$ on $g^{}_{pp}$ is obtained. 
The full-RQRPA has been found to reproduce well the value of  
$M^{2\nu - exp}_{GT}$ deduced from the experimental 
half-live. Agreement is achieved for $^{76}Ge$ and $^{82}Se$
for the parameters $g^{}_{pp} \simeq 0.95$ and $g^{}_{pp} \simeq 1.00$,
respectively. In the case of $2\nu\beta\beta$-decay of $^{128}Te$
and $^{130}Te$ we need higher values of the renormalization parameter
$g^{}_{pp}$: $g^{}_{pp} \simeq 1.3$. We remind that the $Te$ isotopes 
are known to be a very special and delicate case. W. Haxton already
found this out \cite{35}. It could be connected with the fact that the
two Te isotopes 128 and 130, for which the double beta decay is 
measured, are unharmonic or even slightly deformed, so that a 
spherical RPA treatment as used here is not sufficient.

Both, the full-QRPA and full-RQRPA,  include proton-neutron 
pairing correlations via the HFB method. In order to clarify the role 
of proton-neutron pairing in the stabilization of the dependence
of $M^{2\nu}_{GT}$ on $g^{}_{pp}$, we switched off proton-neutron
pairing interaction and performed the calculation within
pn-QRPA and pn-RQRPA (no p-n pairing).  
The results are presented in Figs. 2-5. 
We can see that also in this case one can avoid a strong instability of 
$M^{2\nu}_{GT}$. It is interesting to note that the effect 
due to the proton-neutron pairing is significant. The inclusion 
of proton-neutron pairing in the self-consistent scheme of the
calculation is also helping to shift the instability of $M^{2\nu}_{GT}$ on 
$g^{}_{pp}$ to larger values of $g_{pp}$, 
which are outside the physical range ($0.8 \leq g^{}_{pp} \leq 1.3$).

\section{Conclusions and outlook}

In summary, we have proposed a new higher (renormalized) QRPA
many-body method for nuclear structure studies which we call 
full-RQRPA. We point out that in the full-RQRPA proton-neutron
pairing is taken into account. 
The full-RQRPA includes the Pauli exclusion principle 
in a more proper way in comparison with the QRPA
by using the renormalized quasi-boson approximation. 
The one-particle densities of the correlated state have been 
evaluated by following the approach of Catara et al. 
In this way the ground state correlations beyond 
the RPA are included in the self-consistent scheme of the calculation
consisting of a set of non-linear equations in amplitudes X and Y, which
are solved by iteration.    

We have employed the full-RQRPA to calculate $2\nu\beta\beta$-decay 
of $^{76}Ge$, $^{82}Se$, $^{128}Te$ and $^{130}Te$. The nuclear 
matrix element $M^{2\nu}_{GT}$ governing this process has been 
calculated. We have found $M^{2\nu}_{GT}$ to be rather stable in the
physically acceptable region of the parameter $g^{}_{pp}$, which determine
the strength of the particle-particle interaction. It allows us to 
predict  more reliable values of the nuclear matrix elements. 
The relative large differences between the results of the full-RQRPA and
the full-QRPA are related to violations of the Pauli
exclusion principle by the second method due to 
ground state correlations. 
The quasi-boson approximation is a poor 
approximation to study the $2\nu\beta\beta$-decay. Therefore, 
an alternative methods should be elaborated in which the 
Pauli exclusion principle is explicitly incorporated. 
The shell model approach, which does not violate the Pauli exclusion
principle, fails to construct all needed
states of the intermediate nucleus. We note that the recently
proposed two-vacua random phase approximation method could be perhaps a 
step in the right direction \cite{36,37}. 

The renormalized QRPA which includes the Pauli effect of fermion pairs goes
beyond the quasi-boson approximation. The inclusion of the Pauli principle
reduces the ground state correlations and moves the collapse outside the 
physical region. Our matrix elements are derived from the Bonn-potential
and if we would stick to them, $g_{pp}$ would be unity. Each RPA solution
is collapsing eventually if the forces are made stronger and stronger. The
point is that this  collapse happens here only outside the physical region
as in all stable RPA solutions and therefore the collapse disappeared for the 
description of the nuclei.

The full-RQRPA could find a wider use in the studies of other 
nuclear processes, e.g., single beta decay, neutrinoless double 
beta decay and pion double charge exchange reactions.

\begin{figure}[t]
\vbox{
\vspace*{3.5cm}
\centerline{\epsfig{file=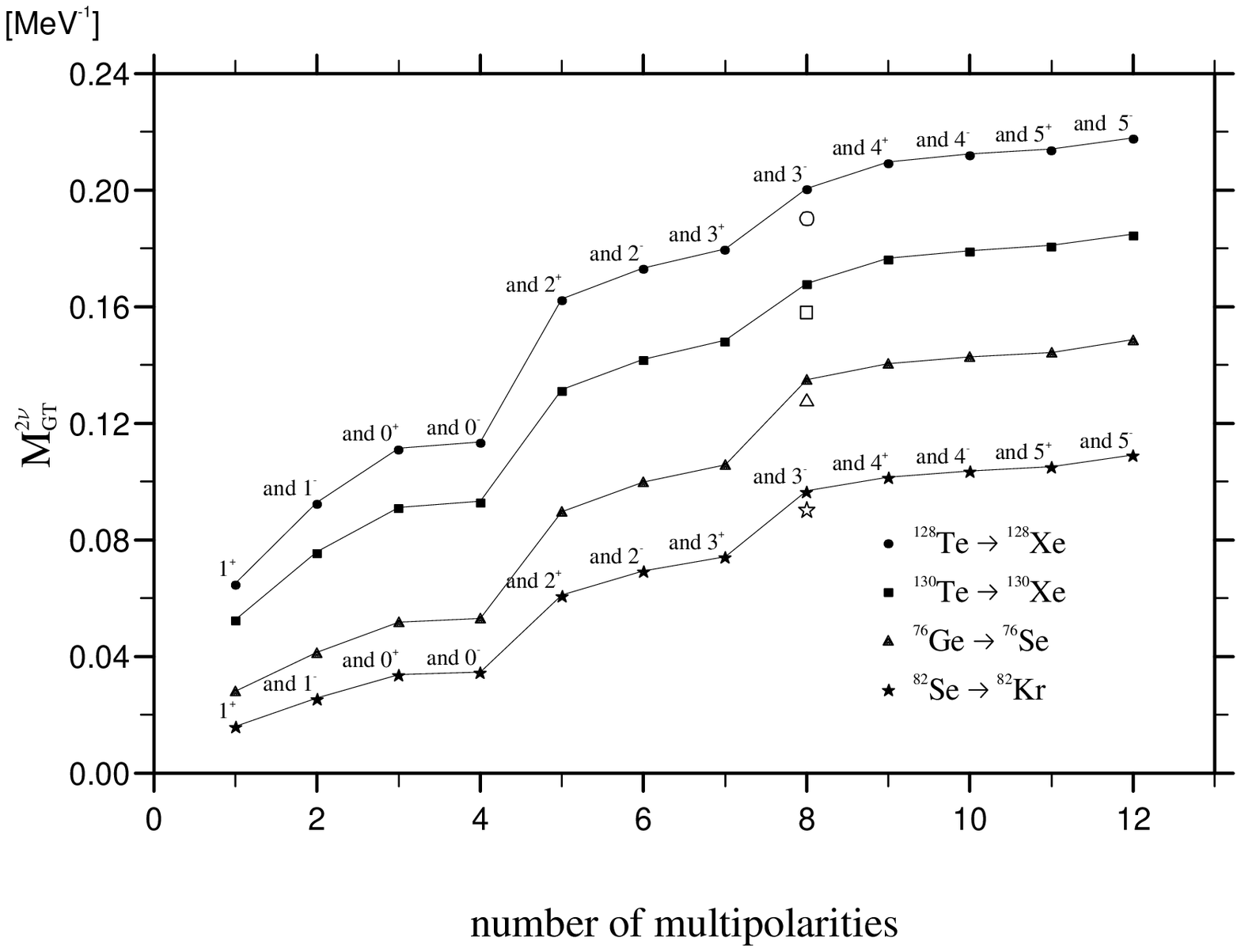,height=4.8in}}
\vspace*{0.9cm}}
\caption{\label{fig:1}
The influence of including more and more multipolarities $J^{\pi}$ 
in the full-RQRPA calculation of $M^{2\nu}_{GT}$ for $g^{}_{pp}=1.0$.
Starting from $1^+$ the mentioned multipole is added to the former ones 
at every point. By this procedure the last point shows the value of
$M^{2\nu}_{GT}$ obtained by the consideration of 
 12 multipolarities ranging from $0^+$ to $5^-$.
The empty symbols mark the values for the combination of 
$1^+$, $2^+$, $2^-$ and $3^-$ multipolarities, which 
turn out to exhaust approximately the most of the correlations beyond RPA.
}
\end{figure}

\begin{figure}[p]
\vbox{
\vspace*{3.0cm}
\centerline{\epsfig{file=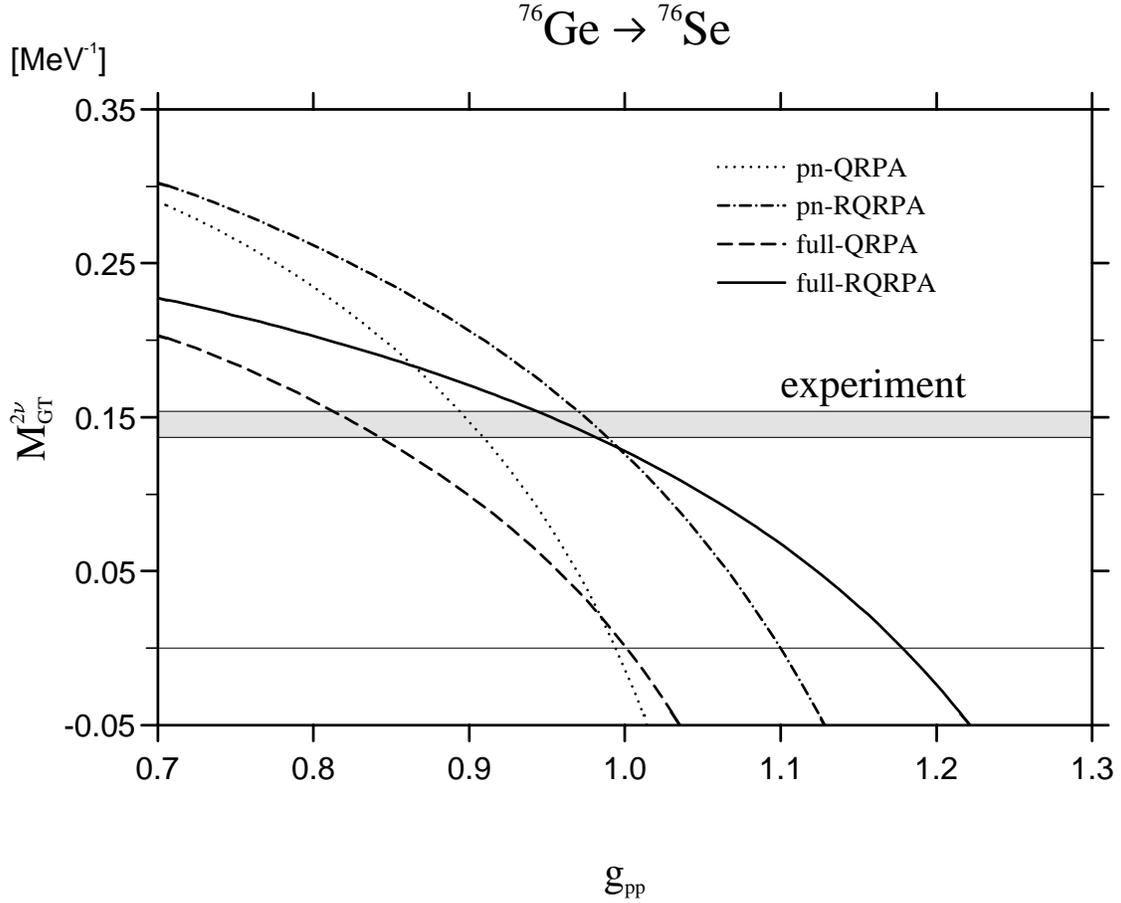,height=4.8in}}
\vspace*{1.0cm}}
\caption{\label{fig:2}
The Gamow-Teller transition matrix element $M^{2\nu}_{GT}$ 
of the $2\nu\beta\beta$-decay of $^{76}Ge$ is plotted as function of 
particle-particle coupling constant $g^{}_{pp}$. The solid line
corresponds to full-RQRPA (with p-n pairing), the dashed line to full-QRPA
(with p-n pairing), the dash-dotted line
to pn-RQRPA (without p-n pairing) 
and the dotted line to pn-QRPA calculation
(without p-n pairing), respectively. }
\end{figure}    

\begin{figure}[p]
\vbox{
\vspace*{3.0cm}
\centerline{\epsfig{file=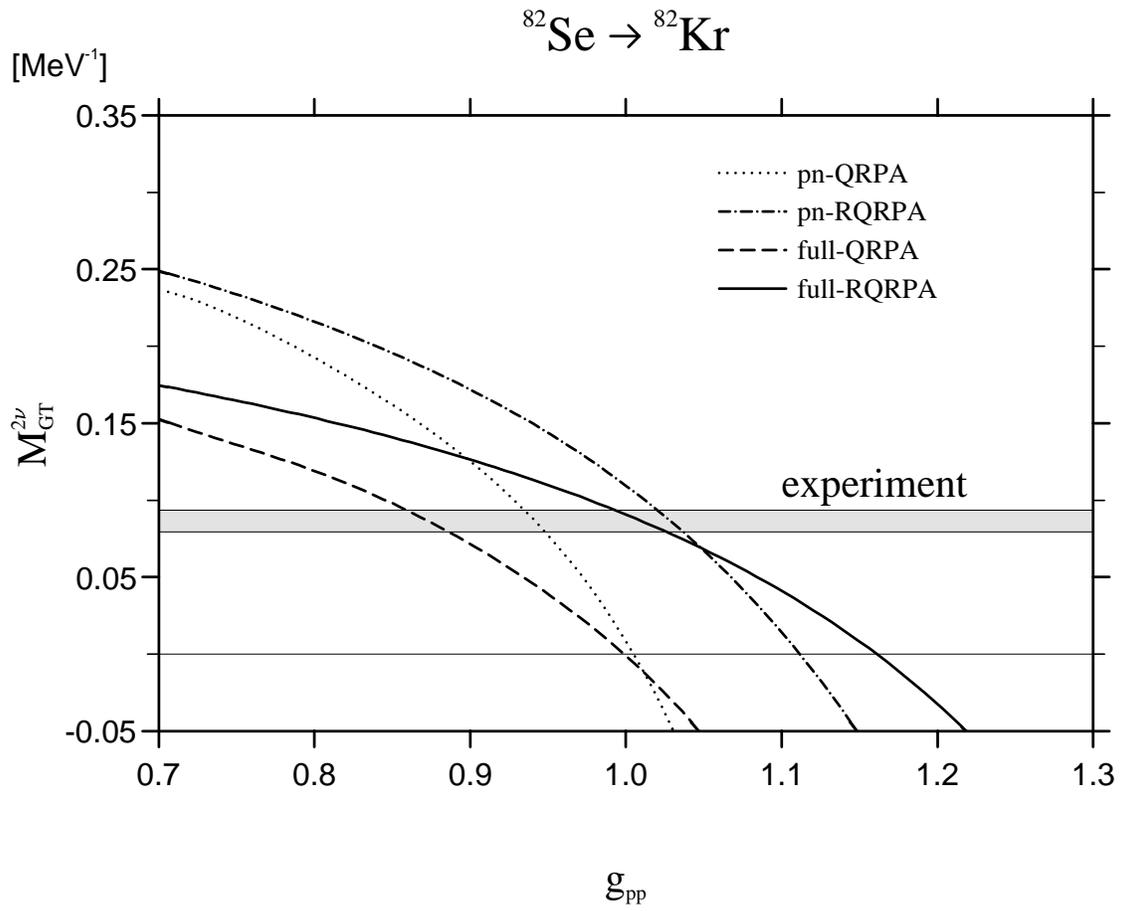,height=4.8in}}
\vspace*{1.0cm}}
\caption{\label{fig:3}
The same as Fig. 2 for $^{82}Se$. }
\end{figure}    

\begin{figure}[p]
\vbox{
\vspace*{-4.00cm}
\centerline{\epsfig{file=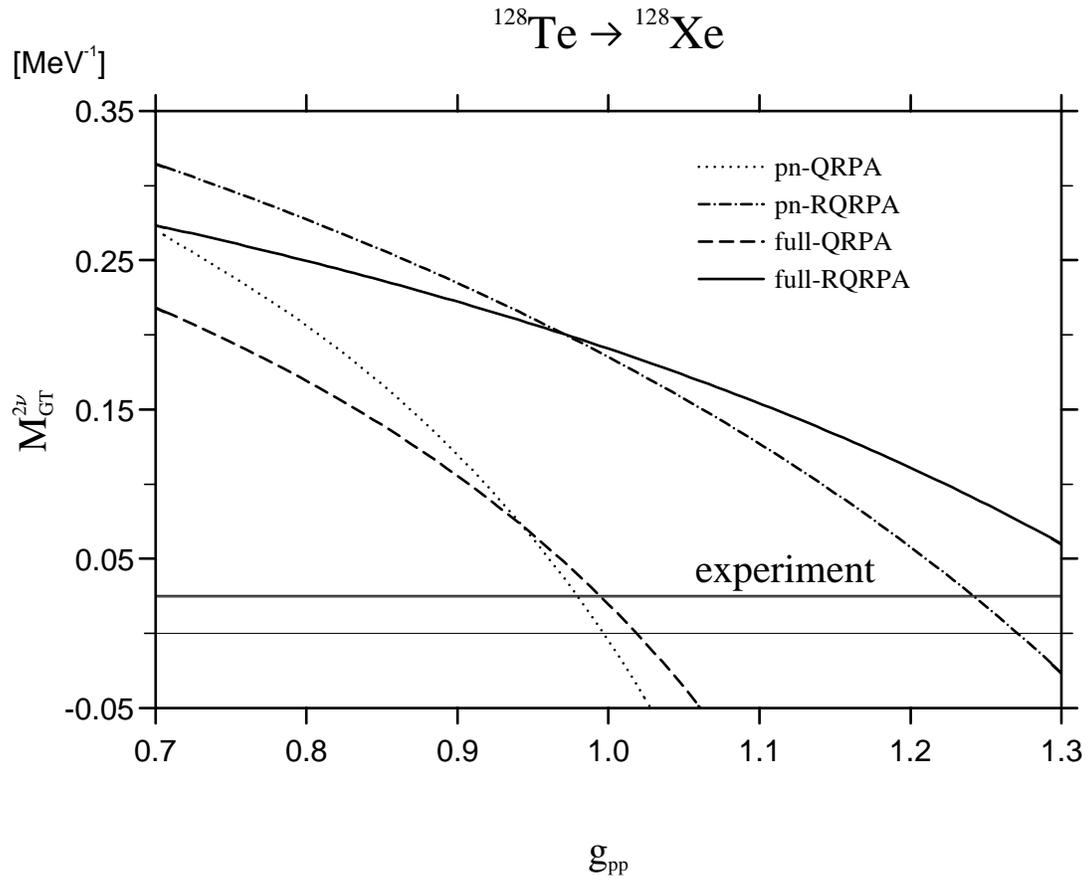,height=8.4in}}
\vspace*{-1.2cm}}
\caption{\label{fig:4}
The same as Fig. 2 for $^{128}Te$. }
\end{figure}    

\begin{figure}[b]
\vbox{
\vspace*{3.0cm}
\centerline{\epsfig{file=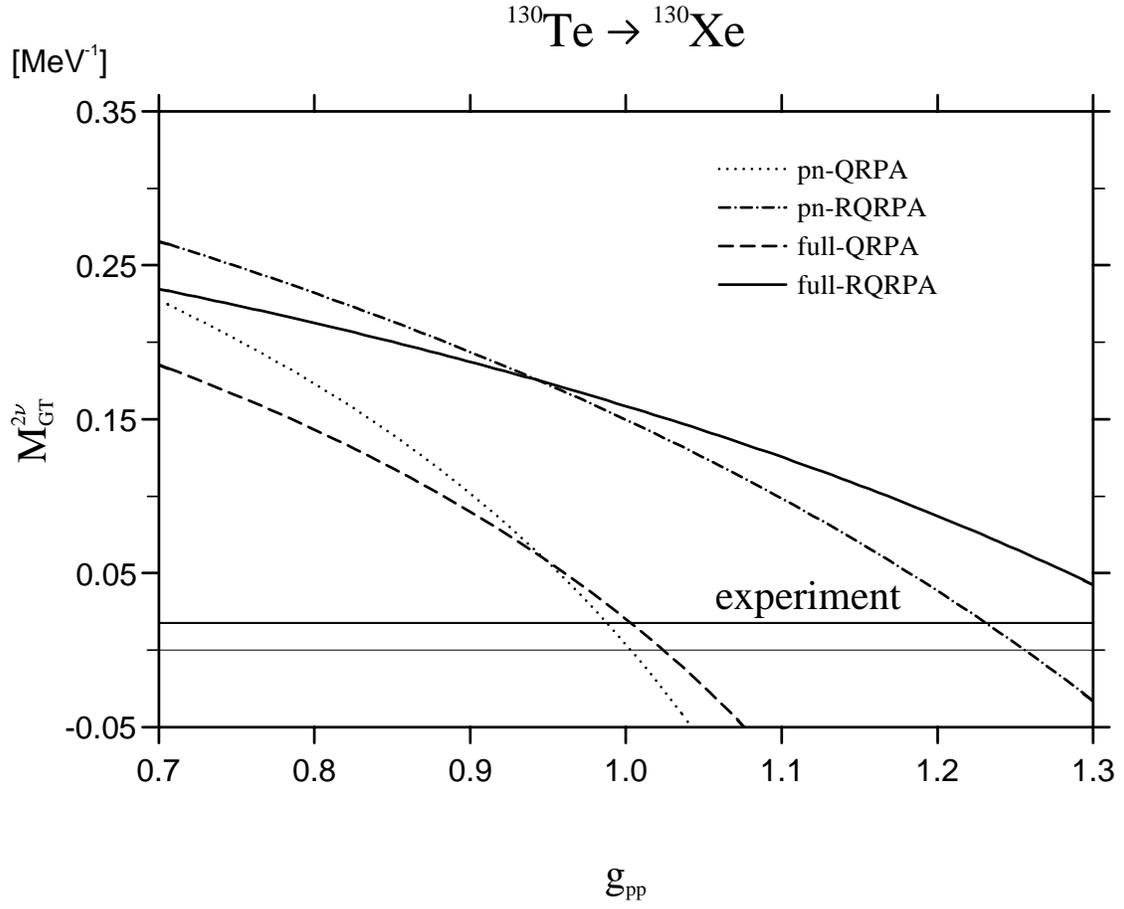,height=4.8in}}
\vspace*{1.0cm}}
\caption{\label{fig:5}
The same as Fig. 2 for $^{130}Te$. }
\end{figure}

\end{document}